# Superior thermal conductivity in suspended bilayer hexagonal boron nitride

Chengru Wang[1], Jie Guo[1], Lan Dong[1], Adili Aiyiti[1], Xiangfan Xu[1,2,†], Baowen Li[3,‡]

[1]Center for Phononics and Thermal Energy Science, School of Physics Science and Engineering, Tongji University, 200092 Shanghai, China

[2]Institute of Advanced Studies, Tongji University, 200092 Shanghai, China

[3]Department of Mechanical Engineering, University of Colorado Boulder, CO 80309-0427, USA

Correspondence and requests for materials should be addressed to
X.X.(xuxiangfan@tongji.edu.cn) or B.L. (Baowen.Li@Colorado.edu)

**We reported the basal-plane thermal conductivity in exfoliated bilayer hexagonal boron nitride *h*-BN that was measured using suspended prepatterned microstructures. The *h*-BN sample suitable for thermal measurements was fabricated by dry-transfer method, whose sample quality, due to less polymer residues on surfaces, is believed to be superior to that of PMMA-mediated samples. The measured room temperature thermal conductivity is around 484 Wm$^{-1}$K$^{-1}$(+141 Wm$^{-1}$K$^{-1}$/ -24 Wm$^{-1}$K$^{-1}$) which exceeds that in bulk *h*-BN, providing experimental observation of the thickness-dependent thermal conductivity in suspended few-layer *h*-BN.**

Hexagonal boron nitride (*h*-BN), analogous to graphene, is an one-atomic layer two dimensional (2D) material with honeycomb structures in which equal Boron and Nitrogen atoms bond compartmentally by sp2 hybridization[1]. Due to their strong covalent bond between B-N and C-C atoms, *h*-BN and graphene hold similar structural and physical properties such as strong mechanical properties, high thermal stability and superior thermal conductivity[2-10]. Consequently, *h*-BN has been proposed to be potential as insulating and dielectric layer for graphene based electronics. Given their geometric similarity and atomic flat on surface, the carrier mobility in graphene/*h*-BN devices has been significantly improved by a factor of 20 at room temperature when comparing to that of graphene on amorphous SiO$_2$ substrate and reaches an ultra high value of 1,000,000 cm$^2$V$^{-1}$s$^{-1}$ at low temperature[11-13]. Furthermore, it's important to note that thermal conductivity in bulk *h*-BN has been found to be as high as 390 Wm$^{-1}$K$^{-1}$, indicating a potential 2D material for efficient heat removal and conduction in further integration and miniaturization of the modern electronics[14].

Several theoretical calculations have suggested that room temperature thermal conductivity in single layer *h*-BN reaches 600Wm$^{-1}$K$^{-1}$ when considering the exact numerical solution of the Boltzmann transport equation, far exceeding that in bulk *h*-BN (~390 Wm$^{-1}$K$^{-1}$)[14-17]. This is reasonable as out-of-plane acoustic phonons are suppressed due to interlayer interaction, as already observed in graphene and graphite theoretically and experimentally[9]. However, recent experiments show that the highest thermal conductivity obtained is around 243 Wm$^{-1}$K$^{-1}$ in 9-layer *h*-BN by Raman method[6] and around 360 Wm$^{-1}$K$^{-1}$ in 11-layer *h*-BN by microbridge device with built-in thermometers[4], respectively. This is understandable that organic residues (e.g. PMMA) and functional groups which are induced during PMMA-mediated wet-transfer process dominate thermal conduction in few-layer BN and reduce its value to below that of bulk *h*-BN[4,18]. Therefore, to study the intrinsic thermal conduction behavior in few-layer *h*-BN, a new transfer technique should be introduced to obtain high quality sample with as less residues as possible.

Here we reported thermal conductivity measurement on suspended bilayer *h*-BN by using prepatterned microstructures with built-in platinum-resistive thermometers. A PMMA-free technique was used to fabricate suspended device suitable for thermal conduction measurement. This dry-transfer method guarantees cleaner surfaces of *h*-BN sample than that in PMMA-mediated method. Consequently, the measured room temperature thermal conductivity reaches a high value of 484 Wm$^{-1}$K$^{-1}$(+141 Wm$^{-1}$K$^{-1}$/-24 Wm$^{-1}$K$^{-1}$), exceeding that in bulk hexagonal boron nitride, indicating *h*-BN as a potential 2D material for efficient heat removal and thermal management in integrated electronic circuit with further miniaturization.

**Results**

We employed the standard prepatterned microstructures for thermal measurements[2,19-21], in which the two Pt/SiNx membranes, named Heater and Sensor in Figure 1b, and their six supporting Pt/SiNx beams were released from silicon substrate by wet etching for 2.5h to 3h. The suspended prepatterned microstructure device was first placed in $O_2$ plasma for 5 min to clean possible organic residue on top of Pt/SiNx membranes. At the mean time, few-layer *h*-BN was exfoliated from BN power by scotch tape method onto PDMS film, in which *h*-BN can been easily identified by different contrast under optical microscope (Figure 1a). Subsequently, *h*-BN/PDMS was aligned under micro-manipulator upside down and attached onto the center of prepatterned microstructure. Due to the strong van der waals force/static electric force, the few-layer *h*-BN was left on Pt electrode after PDMS being peeled off. It is important to note this transfer process is challenging and the sample yield is only few hundred percents. Fortunately, a rectangular *h*-BN between two suspended membranes can be seen in Figure 1b &1c. The device was annealed at 225℃ in $H_2$/Ar atmosphere for two hours to clean the possible residue on top and bottom sides of *h*-BN thin film before any thermal measurements.

Raman spectroscopy is one of the most powerful and fastest methods to detect the number of layers in 2D materials such as graphene and $MoS_2$[22-25]. However, the Stokes shift of Raman peaks varies weekly between different *h*-BN layers, making it impossible to detect the thickness of *h*-BN samples by Raman method[26]. Hence, Atomic Force Microscope was used and a clear step of around 1nm was obtained on edge, indicating two atomic layers in the fabricated *h*-BN sample (Figure 1d).

We followed the same approach employed by Xu *et al*. to measure the thermal conductivity of the suspended device [2]. The device was loaded in Variable Temperature Instrument with vacuum better than $1 \times 10^{-4}$ pa and an *in-situ* anneal was carried out at $T = 450K$ for 2 hours to remove possible residual gas and water molecular. The total measured thermal conductance of *h*-BN sample ($\sigma_s$) and the six supporting Pt/SiNx beams ($\sigma_b$) follows[21]:

$$\sigma_b = \frac{Q_h + Q_s}{\Delta T_h + \Delta T_s}$$

(1)

$$\sigma_s = \sigma_b \frac{\Delta T_s}{\Delta T_h - \Delta T_s}$$

(2)

where $Q_h$ and $Q_s$ are the Joule heating power on the Heater and one supporting

Pt/SiNx beam, respectively. $\Delta T_h$ and $\Delta T_s$, corresponding to resistance changes $\Delta R_h$ and $\Delta R_s$, are the temperature rise on Heater and Sensor, respectively. In previous studies, a uniform temperature distribution was assumed and the average temperature rise $\Delta T_h$ ($\Delta T_s$) in Heater (Sensor) was used instead of the real temperature rise $\Delta T_{R,h}$ ($\Delta T_{R,s}$) at the joint part of sample and Heater (Sensor). This assumption is valid when thermal conductance is low, e.g. $\sigma_s < 0.1\sigma_b$. However for high thermal conductance sample when $\Delta T_h$ is comparable to $\Delta T_s$, according to equation (2), a tiny inaccuracy in measuring $\Delta T_h$ ($\Delta T_s$) will result in significant changes in final $\sigma_s$.

As such, the Finite Element Simulations (COMSOL Multiphysics 5.2, License No: 9400382) was carried out to simulate the temperature distribution in the suspended membranes, Heater and Sensor, at each temperature we measured. Figure 2 shows the simulating results at $T = 300K$. In this simulation, thermal conductivity of Pt can be obtained from the Weidemann-Franz law, thermal conductance of SiNx beam was determined by subtracting the thermal conductance contribution of Pt from the measured Pt/SiN$_x$ beams $\sigma_b$ and its thermal conductivity is calculated to be around 5.8 Wm$^{-1}$K$^{-1}$. The $\Delta T_{R,h}$ ($\Delta T_{R,s}$) was obtained by adjusting the averaged temperature of Heater (Sensor) membrane bounded by dash-dot rectangle to be consistent with the measured temperature rise $\Delta T_h$ ($\Delta T_s$). Figure 2b illustrates the temperature profile cross the platform with $\Delta T_{R,s}$ determined to be 5.344K, comparing to $\Delta T_h$= 6.822K and $\Delta T_s$ = 5.245K. The final thermal conductivity of $h$-BN sample varies 8% and 7.2% at $T = 300K$ and $T = 80K$ respectively after the Finite Element Simulations.

Generally speaking, the total measured thermal resistance $R=1/\sigma_s$ consistent of the contributions from suspended region of $h$-BN sample, $R_{BN}$, and the contacting area between $h$-BN sample and Pt electrode, $R_c$, i.e. $R = R_{BN} + 2R_c$. The thermal contact resistance $R_c$ can be calculated using interfacial thermal resistance ($R_{int}$). The two have been shown to be related as[18]

$$R_c = [\sqrt{\frac{\kappa_c A w}{R_{int}}} \tanh(\sqrt{\frac{w}{\kappa_c A R_{int}}} l_c)]^{-1} \qquad (3)$$

where $\kappa_c$ is thermal conductivity of supported bilayer h-BN (we assume $\kappa_c$ equal to $\kappa$ in suspended bilayer h-BN), $A$ is the cross section area between h-BN sample and electrode, $w$ is the sample width, $l_c$ is the contact length, $R_{int}$ is the interfacial thermal resistance per unit area between BN and Pt electrode. We have not found $R_{int}$ of clean h-BN and Pt interface and the samples in reference have PMMA residues on bottom and top h-BN surface [4], therefore $R_{int}$ of clean graphene and metal interface was used here. Based on the $R_{int}$ data on available literature for graphene measured by the same prepatterned microstructure method, we used the data of $R_{int}$ from the length dependent measurement of suspended single layer graphene with two ends encased by Au and Pt electrode [2]. If this $R_{int}$ is used, the obtained $2R_c/R_{BN}$ ratio reaches 29.1% and 34.1% at $T = 300K$ and $T = 60K$, respectively.

Figure 3a shows the final thermal conductivity $\kappa$ of biyaler h-BN with respect to temperature. $\kappa$ is calculated from $\kappa=\sigma l/(wh)$, where $l = 3\mu m$ is the suspended length, $w = 3.3\mu m$ is suspended width, $h = 0.666nm$ is the thickness of bilayer h-BN as suggested[14,27], $\sigma$ is the obtained thermal conductance after the Finite Element Simulations. The plus error bars (i.e. 29.1% at $T = 300K$ and 34.1% at $T = 60K$) are resulted from the uncertainty in determining the contact resistance $2R_c$ as mentioned above in this manuscript. The measured thermal conductivity increases with temperature and shows a broad plateau/peak when $T > 250K$ due to the Umklapp phonon scattering process dominating the thermal conduction at higher temperature. The plateau/peak value reaches as high as 484 $Wm^{-1}K^{-1}$(+141 $Wm^{-1}K^{-1}$/-24 $Wm^{-1}K^{-1}$), exceeding that in bulk h-BN [14,28]. Meanwhile, the plateau/peak at higher temperature when comparing with that in bulk h-BN single crystal suggests that the extrinsic scattering such as contact, defect and grain size dominance the intrinsic phonon-phonon scattering process at higher temperature. At lower temperature, the thermal conductivity decreases rapidly with its value below that of bulk h-BN when $T < 200K$. This is probably due to the contact [29], relatively smaller grain size [28], or possible tiny residue [4] but non-negligible effect on thermal conductivity.

Nevertheless, the obtained thermal conductivity is larger than that in 5-layer and 11-layer exfoliated *h*-BN measured by modified prepatterned microstructure method (Figure 3a), and that in 9-layer CVD *h*-BN measured by Raman method, suggesting that sample prepared by dry-transfer method holds cleaner surfaces and superior sample quality when comparing to that prepared by PMMA-mediated transfer method[4,6,18].

**Discussion**

Thickness-dependent behavior on basal-plane thermal conductivity of 2D materials is an important topic in thermal transport properties in low dimensional materials and gained intense attractions in last decade. Both theory and experiment suggested that basal-plane thermal conductivity in clean few-layer graphene decreases with increasing layers due to the enhancement of phonon scattering between layers[30-33]. On the other hand, the question on thickness-dependent basal-plane thermal conductivity in $MoS_2$ is still under debate[34,35], as $MoS_2$ has much stronger bond between different layers. Due to the geometric similarity, few-layer *h*-BN has been suggested to have the same thickness dependence similar to graphene[16], yet not observed experimentally. Figure 3b shows the thermal conductivity with respect to layers. At $T$ = 300K, the observed thermal conductivity is larger than that in bulk *h*-BN but smaller than that in single layer *h*-BN by theoretical calculation, indicating a thickness dependent thermal conductivity in few-layer *h*-BN.

It is importing to note that thermal conductivity of supported or encased few-layer graphene and *h*-BN decrease with decreased thickness with the value below that of its bulk counterpart [36]. These two different trends with respect to thickness is understandable as the interaction between graphene (*h*-BN) and substrate materials can also enhance the phonon scatterings in the graphene (*h*-BN) layers. Interestingly, the polymer residues on graphene (*h*-BN) surfaces can also increase the phonon scattering, resulting in opposite trend of the thickness dependent thermal conductivity. In Figure 3b, the previously measured thermal conductivity in 5-layer and 11-layer

*h*-BN with PMMA residue has lower value than that in bulk *h*-BN. It is worth noting that directly comparison of the results from this study and reference [4] is unfair due to the length-dependent thermal conductivity in two-dimensional materials, which has been predicted theoretically [33,37-40] and later confirmed by experiment[2]. However, when comparing the result of sample in this study (with length of 3μm and width of 3.3μm) and that of 12-layers sample in reference [4] (with length of 3μm and width of 9μm), the thermal conductivity in former sample is much larger than that in latter sample with polymer residues on surface, no mention about the slightly width dependent [2,41]. This result provides further evidence that the sample prepared by dry-transfer method has much cleaner surfaces than that obtained by PMMA-mediated method.

In summary, we observed a thickness dependent thermal conductivity in bilayer *h*-BN with the room temperature value reach as high as 484 $Wm^{-1}K^{-1}$(+141 $Wm^{-1}K^{-1}$/-24 $Wm^{-1}K^{-1}$), exceeding that in bulk *h*-BN. Our thermal conduction measurement indicates that the PMMA-free dry-transfer method preserves relatively higher sample quality with less residues on surfaces, providing a brand-new and reliable technique for transferring 2D materials onto suspended prepatterned microstructures suitable for thermal measurements.

## Acknowledgments

This work was supported by National Natural Science Foundation of China (No. 11304227 & No. 11334007) and by the Fundamental Research Funds for the Central Universities (No. 2013KJ024).

## Author contributions

C.W. carried out the sample fabrications and thermal measurements, J.G. contributed to the COMSOL simulations, L.D., A.A. and X.X. helped with the measurement system setup, X.X. wrote the main manuscript text, X.X. and B.L. supervised the project. All the authors contributed to interpret the results and review the manuscript.

# Additional information

Competing financial interests: The authors declare no competing financial interests.

# Figure Legends

**Figure 1. Details of the sample.** (**a**) Optical image of exfoliated bilayer *h*-BN on PDMS. (**b-c**) Optical and Scanning Electron Microscope image of bilayer BN suspending on prepatterned devices; the rectangle in the center which bridging Heater and Sensor is the suspended *h*-BN sample with $l = 3\mu m$ and $w = 3.3~\mu m$. (**d**) Atomic Force Microscope on the edge of *h*-BN sample; the step of ~1 nm is the mark of bilayer *h*-BN. The scale bars are 10 μm.

**Figure 2. Finite Element Simulations (COMSOL Multiphysics 5.2) of the temperature distributions in Sensor.** (**a**) Temperature distribution of Sensor (left) and the whole suspended device (right) at $T = 300K$. (**b**) Temperature profile crossing the Sensor (indicated by dashed line in Figure 2(**a**)) at $T = 300K$.

**Figure 3. Thermal conductivity of the measured bilayer *h*-BN.** (**a**) Thermal conductivity with respect to temperature. The diamonds, pentagons, squares and triangles represent thermal conductivity of single layer *h*-BN (theory) [15], high-quality bulk *h*-BN (experiment) [14], 11-layers suspended *h*-BN and 5-layers suspended *h*-BN (experiment) [4], respectively. (**b**) Layer-dependent thermal conductivity. The left diagonal and right diagonal represent thermal conductivity of high-quality [14] and low-quality [28] bulk *h*-BN (experiment), respectively.

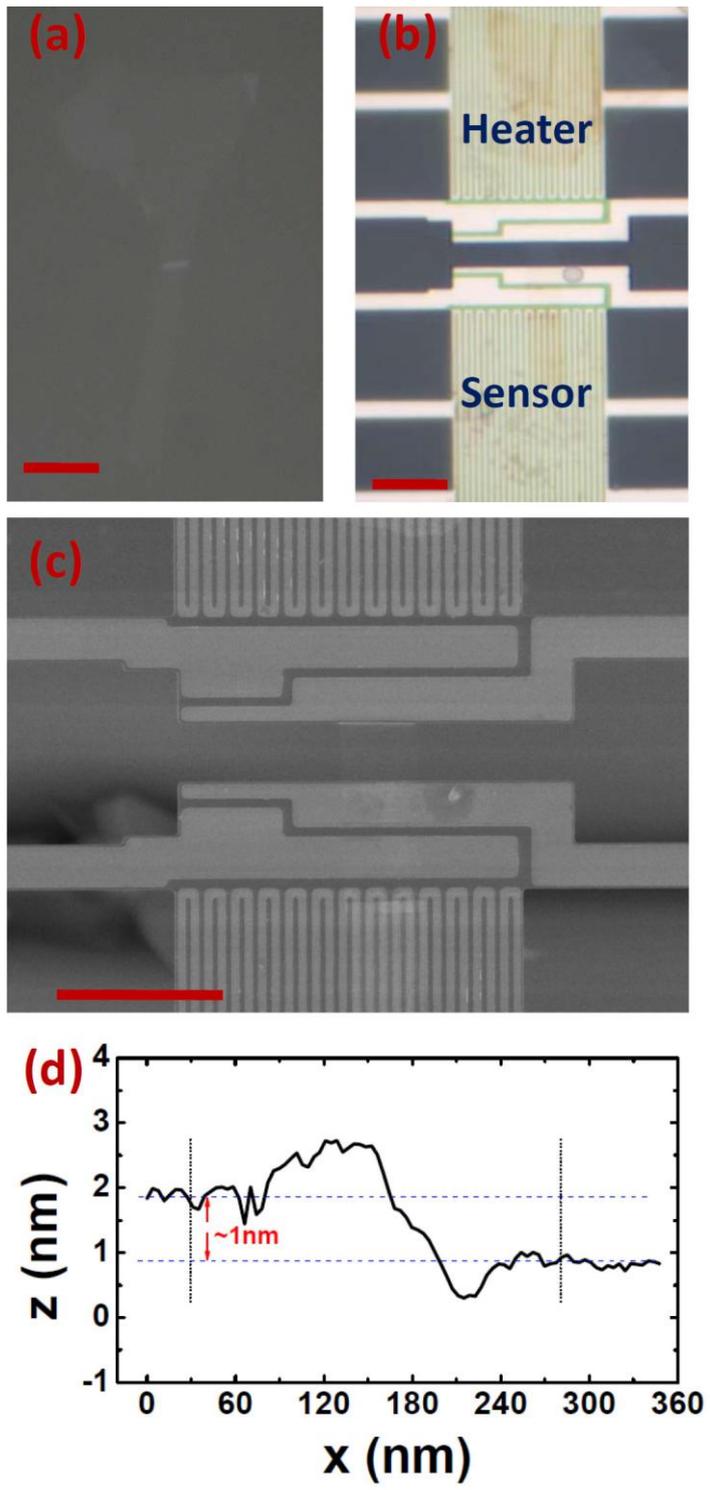

Figure 1.

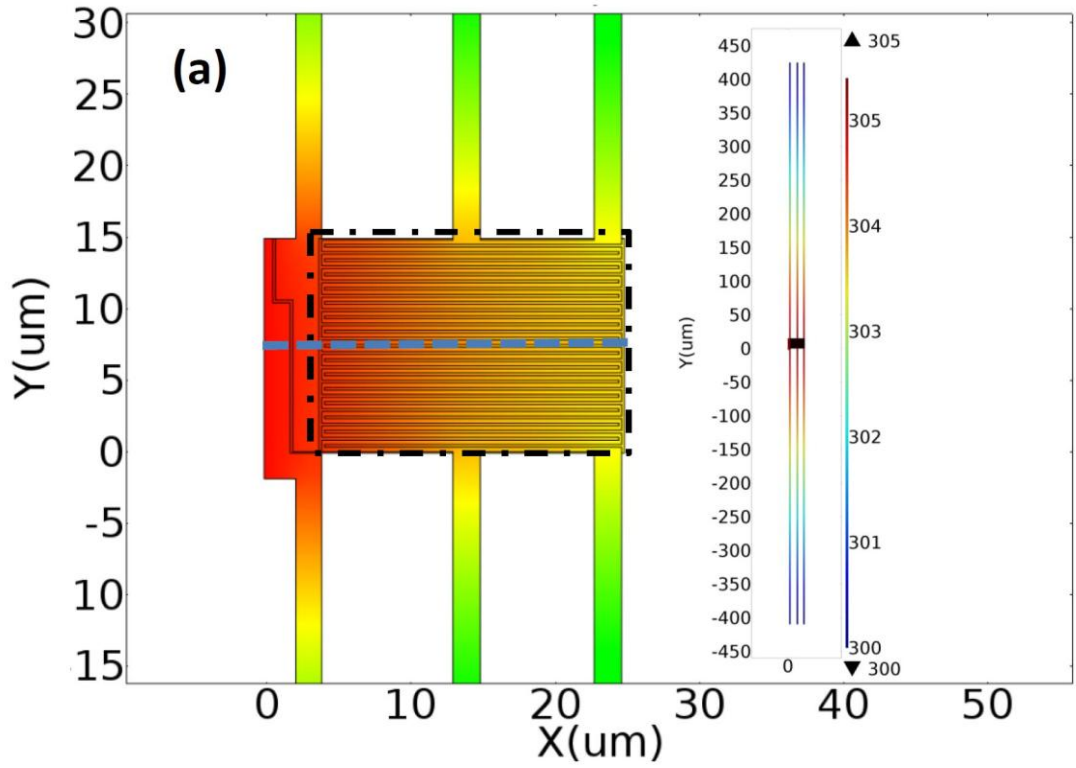

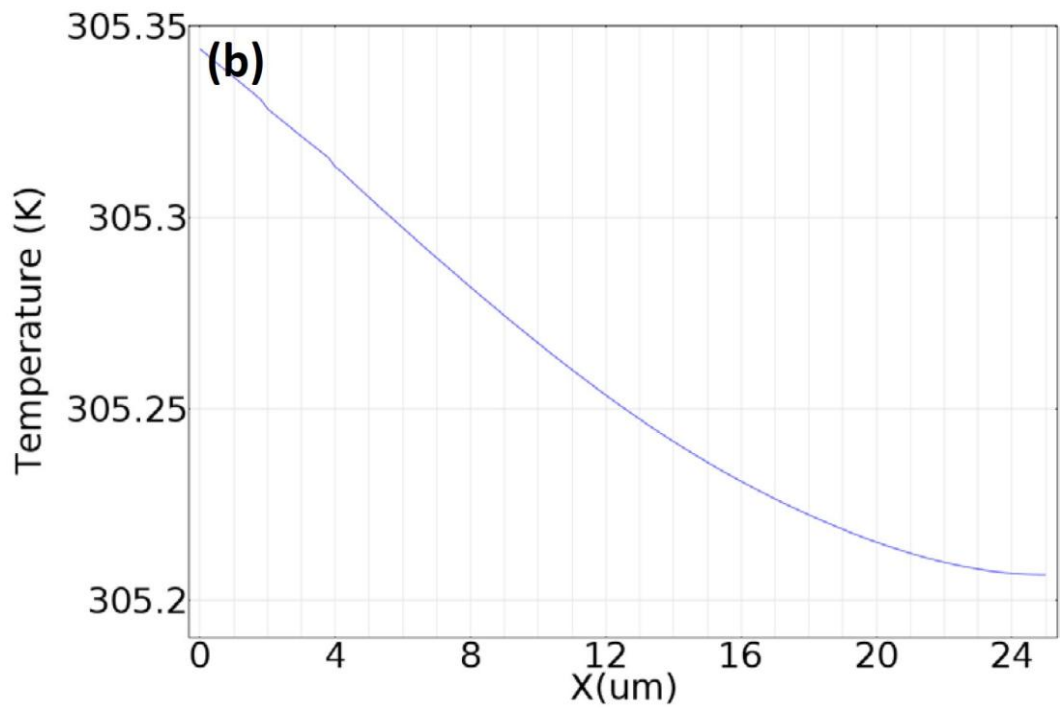

Figure 2.

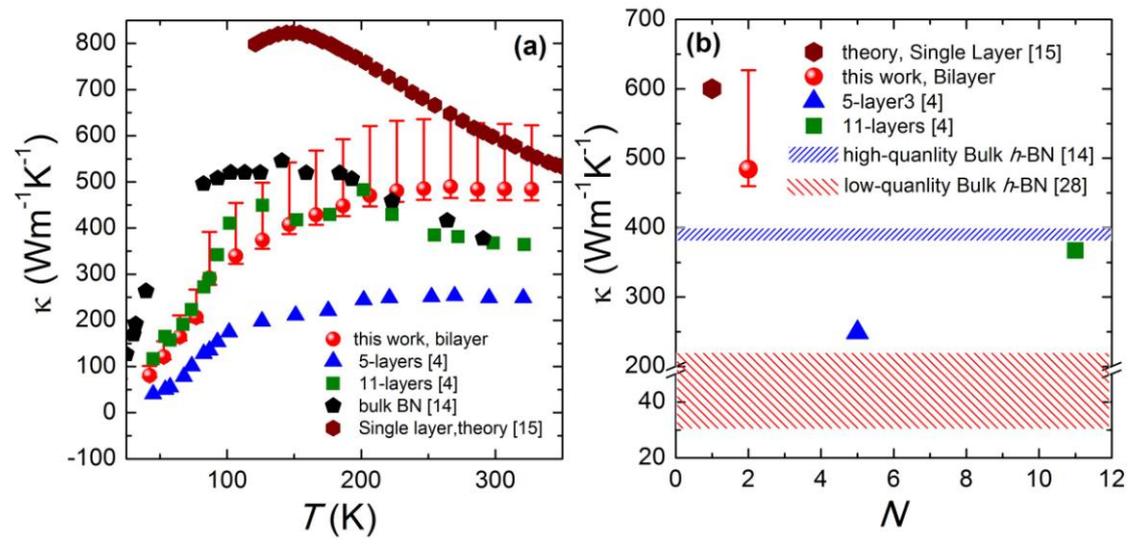

Figure 3